\documentclass[12pt]{iopart}
\usepackage{graphicx}
\usepackage{url}
\usepackage[dvipsnames]{xcolor}
\usepackage{harvard}
\usepackage[normalem]{ulem}
\begin{document}

\title[Double diffraction imaging of nanoparticle dynamics]{Double diffraction imaging of X-ray induced structural dynamics in single free nanoparticles}

\author{M Sauppe$^{1,2}$, T Bischoff$^2$, C Bomme$^{3,4}$, C Bostedt$^{5,6,7}$, A Colombo$^{1}$, B Erk$^{3}$, T Feigl$^{8}$, L Flückiger$^{2,9}$, T Gorkhover$^{2,10}$, A Heilrath$^2$, K Kolatzki$^{1,2}$, Y Kumagai$^{5,11}$, B Langbehn$^2$, J P Mueller$^2$, C Passow$^3$, D Ramm$^3$, D Rolles$^{3,12}$, D Rompotis$^{3,13}$, J Schäfer-Zimmermann$^2$, B Senfftleben$^{2,14,13}$, R Treusch$^{3}$, A Ulmer$^{2,10}$, J Zimbalski$^2$, T Möller$^{2}$ and D Rupp$^{1,2}$}

\address{$^1$ Laboratory for Solid State Physics, ETH Zurich, 8093 Zurich, Switzerland}
\address{$^2$ Institut für Optik und Atomare Physik, Technische Universität Berlin, 10623 Berlin, Germany}
\address{$^3$ Deutsches Elektronen-Synchrotron DESY, 22607 Hamburg, Germany}
\address{$^4$ Saclay Institute of Matter and Radiation, Atomic Energy and Alternative Energies Commission, Gif-sur-Yvette cedex F-91191, France}
\address{$^5$ Chemical Sciences and Engineering Division, Argonne National Laboratory, Illinois 60439, USA}
\address{$^6$ Laboratory for Femtochemistry, Paul Scherrer Institut, 5232 Villigen PSI, Switzerland}
\address{$^7$ Institute of Chemical Sciences \& Engineering, Ecole Polytechnique Fédérale de Lausanne, 1015 Lausanne, Switzerland}
\address{$^8$ optiX fab GmbH, 07745 Jena, Germany}
\address{$^9$ Australian Research Council Centre of Excellence in Advanced Molecular Imaging, Department of Chemistry and Physics, La Trobe Institute for Molecular Science, La Trobe University, Melbourne, Victoria 3086, Australia}
\address{$^{10}$ Institute of Experimental Physics, Universität Hamburg, 22761 Hamburg, Germany}
\address{$^{11}$ Department of Physics, Nara Women’s University, Nara 630-8506, Japan}
\address{$^{12}$ J. R. Macdonald Laboratory, Department of Physics, Kansas State University, Manhattan, KS 66506, USA}
\address{$^{13}$ European XFEL, 22869 Schenefeld, Germany}
\address{$^{14}$ Department of Physics and Astronomy, University of Nebraska - Lincoln, NE 68588, USA}

\eads{\mailto{msauppe@phys.ethz.ch}, \mailto{ruppda@phys.ethz.ch}}

\begin{abstract}
Because of their high photon flux, X-ray free-electron lasers (FEL) allow to resolve the structure of individual nanoparticles via coherent diffractive imaging (CDI) within a single X-ray pulse. Since the inevitable rapid destruction of the sample limits the achievable resolution, a thorough understanding of the spatiotemporal evolution of matter on the nanoscale following the irradiation is crucial. We present a technique to track X-ray induced structural changes in time and space by recording two consecutive diffraction patterns of the same single, free-flying nanoparticle, acquired separately on two large-area detectors opposite to each other, thus examining both the initial and evolved particle structure. We demonstrate the method at the extreme ultraviolet (XUV) and soft X-ray Free-electron LASer in Hamburg (FLASH), investigating xenon clusters as model systems. By splitting a single XUV pulse, two diffraction patterns from the same particle can be obtained. For focus intensities of about $2\cdot10^{12}$\,W/cm$^2$ we observe still largely intact clusters even at the longest delays of up to 650 picoseconds of the second pulse, indicating that in the highly absorbing systems the damage remains confined to one side of the cluster. Instead, in case of five times higher flux, the diffraction patterns show clear signatures of disintegration, namely increased diameters and density fluctuations in the fragmenting clusters. Future improvements to the accessible range of dynamics and time resolution of the approach are discussed.
\end{abstract}

\vspace{2pc}

\section{Introduction}
\label{sec:sec1}

\begin{figure}[tb]
\centering
\includegraphics[width=1\textwidth]{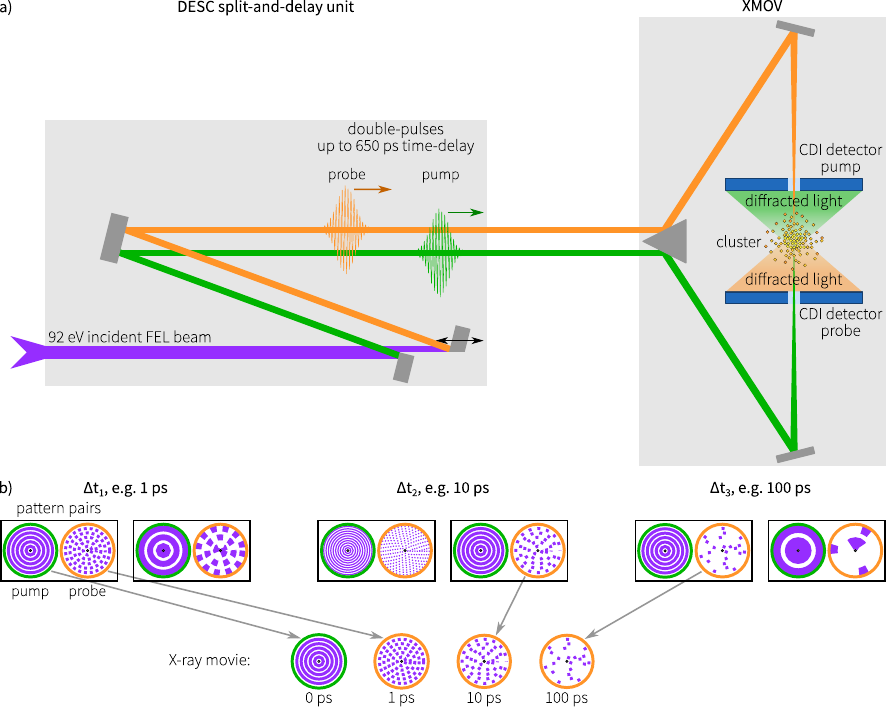}
\caption{Concept of the XMOV. a) Experiment: The pump and probe pulses of 13.5\,nm wavelength (92\,eV) are prepared by the DESC split-and-delay unit from the CAMP end-station \cite{erk2018}. In the XMOV, they are guided onto counter-propagating beampaths. Hence, after the focused beams interact with a cluster, perpendicularly introduced from the top (out of plane), diffracted light from the pump and probe pulse can be separated onto two detectors. b) Post-Processing: Grouping scattering pattern pairs, consisting of a pump and a probe image (here schematics), from different time-delays by matching pump patterns allows to compile their probe patterns to an X-ray movie, see \sref{sec:sec1}. The color code (green = pump, orange = probe) is used throughout the paper.}
\label{fig:fig1}
\end{figure}

Since the discovery of X-rays \cite{rontgen1896} and using them for diffraction on crystals \cite{bragg1913}, structure determination by X-ray diffraction has contributed greatly to a deeper understanding of fundamental mechanisms in nature, as exemplified by the discovery of the DNA double-helix \cite{watson1953} using high-resolution X-ray crystallographs of DNA fibers \cite{franklin1953}. With the method of single-particle coherent diffractive imaging (CDI) \cite{chapman2010,miao2015} using short-wavelength free-electron lasers (FEL) we can even capture the structure of individual free-flying nanoscale objects within a single, intense X-ray pulse. Fragile objects such as single viruses \cite{seibert2011,ekeberg2015}, bacteria \cite{vanderschot2015}, cell organelles \cite{hantke2014}, aerosols \cite{bogan2010a,loh2012}, atomic clusters \cite{bostedt2012,rupp2012,barke2015} and superfluid helium nanodroplets \cite{gomez2014,langbehn2018} have been successfully imaged. Also ultrafast laser-induced dynamics of nanoparticles have been ``filmed'' in pump-probe configurations \cite{gorkhover2016,fluckiger2016,ihm2019a,peltz2022,bacellar2022a,langbehn2022,dold2023}.

In order to produce a sufficiently bright diffraction pattern of a single nanostructure in a single irradiation, the required focal intensities are so high that the particles are rapidly transformed into a highly excited non-equilibrium state, leading to their fast destruction \cite{neutze2000}. The assumption which is made in CDI in this regard is termed ``diffraction before destruction'', signifying that the X-ray pulse has to be sufficiently short so that significant atomic motion sets in only after the pulse \cite{neutze2000,chapman2006}. While this assumption can be valid in many applications using larger particles or samples consisting of heavy materials, ultrafast radiation damage limits the achievable resolution of single-particle CDI, and does so particularly when addressing single bio-molecules consisting of lighter compounds and aiming for time-resolved ``molecular movies'' \cite{neutze2000,hau-riege2004}. Changes of the particle's electronic structure due to ionization and plasma formation occur in parallel to diffraction and cannot be outrun \cite{bostedt2012,rupp2020,ho2020}. Therefore, from the start of X-ray free-electron laser science, substantial efforts were directed towards developing a thorough understanding of the interaction of high-intensity X-rays with matter. Many experimental and theoretical studies have been carried out in this regard using atomic clusters as ideal model systems, being scalable in size and complexity and well accessible for theoretical modelling \cite{bostedt2010,bostedt2016,bostedt2020,young2018,saalmann2006,fennel2010}. A combined approach of CDI and simultaneous ion spectroscopy of single atomic clusters \cite{gorkhover2012,rupp2016,saladrigas2021a} has yielded particularly detailed insights because the size and shape of the individual cluster at the beginning, as well as the FEL's intensity can be determined from each coherent diffraction pattern. These can be used for sorting the respective ion spectra for these quantities. Filming of X-ray induced dynamics in atomic clusters, however, is quite difficult, as dynamics occur on a femto- \cite{gorkhover2016} to nanoseconds \cite{schutte2014a,fluckiger2016} timescale, much faster than the currently fastest CDI detectors, which allow to record consecutive patterns with around 100\,ns separation \cite{allahgholi2019}. Therefore, both the image of the initial particle, as well as a second image of the evolved system have to be captured separately. To this end, we developed an experimental method based on geometrically separating two images produced by two time-delayed counter-propagating FEL pulses using two opposing detectors. The concept termed ``X-ray MOVie camera'' (XMOV) is sketched in \fref{fig:fig1}\,a). While from the CDI pattern generated by the pump pulse, the particle's shape, size, orientation and exposed intensity can be extracted, the pump-pulse-induced structural changes are probed in the second diffraction pattern created by the probe pulse. As the pump pattern provides full information on each individual nanoparticle, also non-reproducible targets can be investigated, but may require longer measurement time to investigate structural effects. \Fref{fig:fig1}\,b) exemplifies how the XMOV allows to follow the dynamics also in a longer sequence. Typically CDI pattern pairs are recorded for several time-delays. If one groups the pairs for a matching starting condition, i.e.\@ by (nearly) identical pump patterns, the respective probe patterns from different time-delays can be compiled to a proper ``X-ray movie''. We note that a related concept was proposed in \cite{schmidt2008} and a counter-propagation setup without CDI detection was realized in \cite{rompotis2017}. Using only a single detector, an ultrafast movie was demonstrated applying holography \cite{gunther2011}. In the hard X-Ray regime, two X-ray pulses were used simultaneously for stereo imaging \cite{hoshino2011} and to obtain 3D-information \cite{bellucci2023}, or concepts proposed for time-resolved crystallography \cite{vanthor2015}. Recently, also two pulse X-ray photon correlation spectroscopy was demonstrated \cite{roseker2018}.

We demonstrated our approach in two experimental campaigns at the CAMP end-station (CFEL-ASG Multi-Purpose instrument) \cite{erk2018} at beamline BL1 of the FLASH (Free-electron LASer in Hamburg) FEL operating in the extreme ultraviolet (XUV) and soft X-ray range \cite{ackermann2007}. This beamline also hosts the split-and-delay unit DESC (DElay Stage for CAMP) providing 13.5\,nm ($\hat{=}$ 91.8\,eV) double pulses with time-delays up to 650\,ps (see left side of \fref{fig:fig1}\,a)) \cite{sauppe2018}. As model systems, xenon clusters with radii in the order of 100\,nm were used. Due to their size and high atomic mass, it is to be expected that no significant atomic motion will occur already during the pump pulse. Further, at 90\,eV photon energy, a broad resonance in xenon \cite{ederer1964,lukirskii1964} results in a strong scattering response. Using XUV pulse pairs with zero time-delay, two CDI patterns of the same single nanostructure in free flight could be detected on a regular basis. This is shown both by events where specific features from twinned clusters were observed on both detectors, and by a statistical analysis of many diffraction pattern pairs, thus serving as proof of concept. However, in both experiments, the quality of the CDI patterns, the achieved statistics, and the accessible dynamics, were limited by rather low focal intensities due to surface errors of the XUV focussing optics and by parasitic stray light. For this reason, the analysis presented in this work has a qualitative character and the interpretation of the dynamics is given in terms of possible physical pictures. In the first experiment, focal intensities in the lower $10^{12}$\,W/cm$^2$ range were reached, limiting the dynamics to the onset of damage. Even at longest time-delays of 650\,ps seemingly unchanged clusters were observed by the probe pulse. We explain this observation with the high absorption of xenon clusters, limiting the ionization by the pump pulse to only a small slice of the cluster. In consequence, only the ``front'' side of the cluster, facing the pump pulse, is structurally affected, leaving behind an apparently unchanged cluster to the opposing probe pulse, impinging the ``rear'' side. At a five times higher XUV flux in the second experiment, fluctuations resulting in speckle patterns are observed, serving as an indication of cluster disintegration.

This paper is structured as follows. \Sref{sec:sec2} describes the XMOV setup and implementation as well as the experimental conditions of both campaigns at FLASH. In \sref{sec:sec3}, the successful double-diffractive imaging of the same single xenon clusters is discussed. The temporal resolution of the current setup, limited to a few picoseconds, is analysed in \sref{sec:sec4}. In \sref{sec:sec5}, the time-resolved results at lower focal intensities in the first experiment are compared to results at higher intensities of the second experiment. Finally, possible advancements of this method are discussed in \sref{sec:sec6}, such as precise tracking of cluster positions and changes in mirror geometries to overcome the limitations to dynamic range and time resolution of this first realization of an X-ray movie camera.

\section{Experimental method}
\label{sec:sec2}

\begin{figure}[tb]
\centering
\includegraphics[width=1.00\textwidth]{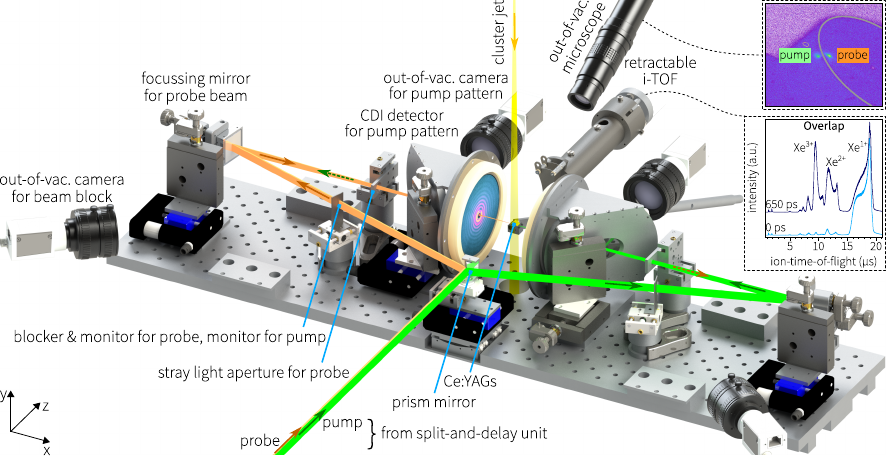}
\caption{Compact in-vacuum optical setup on a breadboard. In this CAD drawing, the Ce:YAG screens are shown inserted into the interaction-region while the ion time-of-flight spectrometer (i-TOF) is retracted. The upper inset illustrates the microscope view of both foci on the edge (gray line) of a Ce:YAG screen. The lower inset shows ion spectra of small xenon clusters (average radius 38.3\,nm) at successfully established spatial overlap for two time-delays (650\,ps shown with offset). Note that the prism mirror is named such due to its wedge shape, it is not a diverging prism.}
\label{fig:fig2}
\end{figure}

\fulltable{\label{tab:tab1} Motorized axes for in-vacuum beam alignment. Given is the respective device with possible motions along or around an axis (compare \fref{fig:fig2}) and the major alignment effects on the beam.}
\br
Device                        & Motion               & Alignment effect on beam                                     \\
\mr
\textbf{prism mirror}         & positioning along x  & adjust pump-probe splitting                                  \\
                              & positioning along z  & position on focussing mirror                                 \\
                              & rotating around y    & angle on prism mirror, angle \& position on focussing mirror \\
                              & rotating around z    & angle on prism mirror, angle \& position on focussing mirror \\
\mr
\textbf{beam block/}          & positioning along z  & block and monitor one arm and monitor other                  \\
\textbf{monitor ($\times2$)}  & (with some angle)    &                                                              \\
\mr
\textbf{focussing}            & positioning along x  & position on focussing mirror \& spatial overlap              \\
\textbf{mirror ($\times2$)}   & positioning along y  & position on focussing mirror \& spatial overlap              \\
                              & positioning along z  & position on focussing mirror \& spatial overlap              \\
                              & pitch                & focus tuning                                                 \\
                              & tilt                 & focus tuning                                                 \\
\mr
\textbf{stray light}          & positioning along z  & aperture centering                                           \\
\textbf{aperture ($\times2$)} & positioning along y  & aperture centering                                           \\
\mr
\textbf{CDI}                  & positioning along z  & detector centering                                           \\
\textbf{detector ($\times2$)} & positioning along y  & detector centering                                           \\
\br
\endfulltable

The undulator system of the FLASH1 branch offers high pulse energies in the XUV regime \cite{ackermann2007}. The CAMP end-station \cite{erk2018} is equipped with the delay stage DESC, which allows to carry out single-color XUV pump-probe experiments up to long time-delays in the hundreds of picosecond range \cite{sauppe2018}, thus making CAMP the ideal choice for demonstrating the XMOV concept.

For the experiments, the end-station's focussing unit \cite{erk2018} had to be retracted and replaced by a beam transport section, and the center of the main chamber of CAMP was aligned to the unfocused beam. Further, CAMP was extended on both sides to fit the optical setup, while the usually applied pnCCD-detectors at the back of CAMP \cite{struder2010} were replaced by a multiport flange for, among others, out-of-vacuum cameras and a manipulator with the ion time-of-flight spectrometer. On top of the interaction chamber, a differentially pumped chamber for the cluster source was mounted. The in-vacuum setup was assembled on a breadboard, which was fixed on mounting points in the CAMP chamber and fits with the full setup through a DN 250 CF flange. This allows for handy pre-alignment of the setup out-of-vacuum. A CAD drawing of the XMOV breadboard with its main components is shown in \fref{fig:fig2}.

The split-and-delay unit DESC is based on multilayer mirrors and readily delivers XUV double pulses at 13.5\,nm wavelength \cite{sauppe2018}. Accordingly, we chose also multilayer mirrors optimized to the same wavelength to steer the XUV pulses, which allowed us a more compact and relaxed geometry as compared to grazing incidence mirrors. At 13.5\,nm wavelength multilayers can be produced with high reflectivities of 63\,\% at near-normal incidence \cite{feigl2006,bourassin-bouchet2015}. The downside of this choice is of course that the wavelength is fixed. Also, because of the FEL's statistical nature of light creation \cite{seddon2017}, shot-to-shot fluctuations in wavelength may influence the transmission. However, the used multilayer mirrors have a bandwidth of 0.5\,nm (FWHM) \cite{sauppe2018} and the FEL was tuned on a stable average transmission via two mirrors.

The simplified beampath through DESC and XMOV is presented in \fref{fig:fig1}a) but the distances and angles are not to scale. In the actual setup (see \fref{fig:fig2}), DESC delivers the double pulses in two spatially separated half-circled beams \cite{sauppe2018}, which arrive in the middle of the XMOV breadboard and are diverted by a prism (i.e. wedged) mirror into opposing directions at an incidence angle of $54^\circ$ to the normal of each prism face. For in-vacuum beam alignment, the setup is equipped with a total of 24 motorized axes, see alignment effects on the beams in \tref{tab:tab1}. Pump and probe beams are then focused by two mirrors at the long ends of the breadboard under an incidence angle of $9^\circ$ to the surface normal in anti-parallel direction (along the x-axis, see coordinate system in \fref{fig:fig2}) through central holes in the CDI detectors into the same spot, the interaction region. For the reduction of parasitic stray light, round apertures with fixed sizes can be inserted. Further, each beam can be separately blocked. Photoluminescent coatings and crosshairs on both sides of the beam blockers allow to use them also for monitoring the in- and outgoing beams with out-of-vacuum cameras. A pair of tilted photoluminescent Ce:YAG screens can be introduced into the interaction region for focus characterization and spatial overlap optimization with few micrometer precision using an out-of-vacuum microscope (see upper inset in \fref{fig:fig2}). Fine focus optimization and final spatial overlap are achieved using a movable and fully retractable ion time-of-flight spectrometer and optimizing on high charge states of a gas or cluster target (see lower inset in \fref{fig:fig2}). The CDI detectors consist of chevron microchannel plates with 75\,mm effective diameter, stacked with a phosphor screen \cite{bostedt2010}. All plates have a central hole to let the XUV beams pass through. Out-of-vacuum cameras capture the CDI patterns which are made visible on the phosphor screens and reflected by plane, center-drilled optical mirrors. The CDI detectors are placed such that the MCPs have a distance of 38\,mm to the interaction region. This geometry allows for recording a maximum scattering angle of $45^\circ$. Temporal gating of the MCP voltages ensures that solely photons and no ions are detected. The cluster jet is introduced to the interaction region from the top (along the y-axis).

Despite the multilayer mirrors' comparably high reflectivity, multiple reflections and an incidence on the prism mirror close to Brewster's angle limit the transmission of the XMOV to 24\,\%. Accounting further for the transmission of the split-and-delay unit and the beamline itself, the calculated overall transmission in 50/50 splitting reduces the beams in each arm to 3\,\% of the full FEL pulse energy as determined by the FLASH gas monitor detector \cite{tiedtke2009}. In order to still achieve bright diffraction patterns, a strong focusing is required. In the first experimental campaign, a pair of ellipsoidal mirrors were used. This set of mirrors had an insufficient substrate surface figure leading to minimum spot sizes of 61\,\textmu m (FWHM, assuming Gaussian profile). For the second campaign, a pair of toroidal mirrors achieved 26\,\textmu m spots (FWHM, assuming Gaussian profile). We note that toroids are easier to produce with sufficient surface figure quality, but they have an inherent aberration (coma), which ultimately limits the achievable peak intensity.

\Table{\label{tab:tab2} Experimental parameters. The wavelength was matched to the multilayer mirrors of the split-and-delay unit DESC and the optical setup of XMOV (identical requirement). The spot size in experiment 1 was calculated using interferometrically measured substrate surface data, while in experiment 2 calibration measurements were carried out and compared with \cite{gerken2014}. The values are given assuming Gaussian focus profiles which is a simplification, in particular for the first experiment. The focal intensities are calculated with the mean pulse energies ($\pm$ standard deviation) taking into account the respective intensity-ratios of pump and probe beam, i.e. the splitting in DESC, which was different for experiment 1 and experiment 2, as well as the transmission for this wavelength of the beamline, DESC, and XMOV.}
\br
Parameter               & Experiment 1                   & Experiment 2                     \\
\mr
Wavelength              & \centre{2}{13.5\,nm ($\hat{=}$ 91.8\,eV)}                         \\
FEL pulse energy        & ($215 \pm 15$)\,\textmu J      & ($220 \pm 25$)\,\textmu J        \\
Pulse duration          & 100\,fs                        & 200\,fs                          \\
Spot size (calculated)  & 61\,\textmu m                  & 26\,\textmu m                    \\
Intensity pump pulse    & $1.6 \cdot 10^{12}$\,W/cm$^2$  & $4.0 \cdot 10^{12}$\,W/cm$^2$    \\
Intensity probe pulse   & $1.9 \cdot 10^{12}$\,W/cm$^2$  & $7.2 \cdot 10^{12}$\,W/cm$^2$    \\
\br
\endTable

With these two sets of XUV focusing mirrors, we performed two XMOV experiments on xenon clusters. The xenon clusters were generated in the post-pulse \cite{rupp2014} of a pulsed supersonic gas expansion leading to a size distribution around 80\,nm in radius with 17\,nm standard deviation. The cluster source was installed on top of the interaction region at a distance of $\approx275$\,mm and differentially pumped to reduce the gas load in the main chamber. The central part of the cluster beam was guided through a 0.5\,mm conical skimmer mounted at a distance of 207\,mm above the interaction region. Further conditions of the two experimental campaigns are provided in \tref{tab:tab2}.

In the first experiment, much time was needed for aligning and optimizing the novel setup, which limited the measurement time to six hours. Further, the low intensities in the interaction region resulted in a low hit rate as only the brightest events caused detectable CDI patterns. Clear features of diffracted light could be observed only up to a maximum scattering angle of $25^\circ$. All \emph{hits}, defined as events which contain patterns from single clusters \emph{on one or both detectors}, are summarized in \tref{tab:tab3}.

\fulltable{\label{tab:tab3} Overview of all hits in the first experiment. The hits' relative abundance is given in parentheses and calculated with respect to single shots. The order of recording the time-delays was 0\,ps, 650\,ps, 70\,ps.}
\br
Beam configuration              & Event type                 & 0\,ps              & 70\,ps         & 650\,ps         \\
\mr
\textbf{pump \& probe pulse}    & shots                      & 26329              & 11483          & 4972            \\
                                & only pump pattern visible  & 19 (0.07\,\%)      & 44 (0.38\,\%)  & 57 (1.15\,\%)   \\
                                & only probe pattern visible & 365 (1.39\,\%)     & 145 (1.26\,\%) & 70 (1.41\,\%)   \\
                                & both patterns visible      & 78 (0.3\,\%)       & 20  (0.17\,\%) & 23 (0.46\,\%)   \\
\mr
\textbf{only pump pulse}        & shots                      & 5663               & 2651           & 2235            \\
                                & pump patterns              & 64 (1.13\,\%)      & 10 (0.38\,\%)  & 22 (0.98\,\%)   \\
\mr
\textbf{only probe pulse}       & shots                      & 4051               & 1487           & 1009            \\
                                & probe patterns             & 141 (3.48\,\%)     & 44 (2.96\,\%)  & 54 (5.35\,\%)   \\
\br
\endfulltable

For each time-delay, by blocking either one or the other arm, also data only with pump pulse or only with probe pulse were taken for cross-checks. With pump and probe pulse present, a total of 121 CDI pattern pairs with both a pump hit and probe hit have been identified for all three time-delays. The majority of all pattern pairs taken in pump-probe configuration and classified as hits, however, show only a pump pattern or a probe pattern, respectively. It is likely that at least at zero delay, these events are caused by an imperfect spatial overlap or spatial jittering of the focal profiles, such that the respective cluster was only sufficiently irradiated by one of the two pulses to cast a detectable pattern. This is further discussed in \sref{sec:sec5}. As a first step, however, the patterns at zero time-delay showing signal on pump and probe detector can be used to demonstrate that indeed in most events with two patterns, the same particle is imaged twice in the XMOV approach, as will be shown in the following section.

\section{Proof of concept of the XMOV approach}
\label{sec:sec3}

\begin{figure}[t]
\centering
\includegraphics[width=1.00\textwidth]{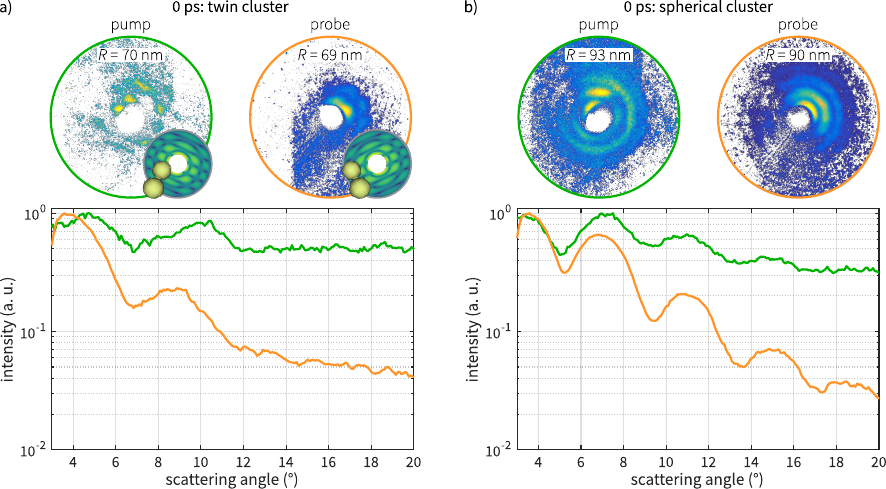}
\caption{CDI pattern pairs and their radial profiles at 0\,ps, shown up to a scattering angle of $20^{\circ}$. a) Twice imaged twin cluster. For clarification simulated scattering patterns of the shown 3D model are inset. b) Twice imaged spherical cluster showing concentric rings with matching spacing. We would like to point out that separate background corrections have been carried out for pump and probe images so that the slopes cannot be compared. In addition, the contrast was increased for the visualization of the patterns and remaining background was removed.}
\label{fig:fig3}
\end{figure}

For a time-resolved analysis it must be first verified that in a CDI pattern pair with pump hit and probe hit both stem from the same particle. Back-scattering can be ruled out as solely forward scattering is detected when one beam is blocked.

The first argument is a statistical one. We note that all statistics in this work must be viewed with caution; they rely on low numbers and depend on many experimental factors, like symmetrical alignment, FEL intensity and pointing fluctuations as well as different stray light backgrounds, which all may lead to a correlated probability of the visibility of hits for pump and probe detector. Nevertheless, hits taken only with the pump pulse or the probe pulse (\tref{tab:tab3}) at least allow to estimate a rough upper bound for the uncorrelated probability to measure by coincidence a CDI pattern pair with both pump pattern and probe pattern stemming from different clusters. On average for all three time delays it follows $0.91\,\% \cdot 3.65\,\% = 0.03\,\%$ (only pump pulse $\cdot$ only probe pulse).\footnote{Explanation for the calculation of the relative abundances, using the values from \tref{tab:tab3}: only pump pulse: $(64+10+22)/(5663+2651+2235) = 0.0091$, only probe pulse: $(141+44+54)/(4051+1487+1009) = 0.0365$.} In contrast, CDI pattern pairs with both pulses present and pump hit and probe hit at 0\,ps have a relative abundance of 0.3\,\%, for 70\,ps 0.17\,\% and for 650\,ps 0.46\,\% (values taken from \tref{tab:tab3}). These larger values, however, suggest that a correlated effect actually plays a role here, and making it statistically likely that most of them actually stem from the same xenon cluster.

A stronger argument can be made by the direct comparison of the structures shown by each pump pattern and probe pattern at 0\,ps. A gas expansion for cluster production always leads to a size distribution \cite{pocsik1991} and, under the expansion conditions chosen in our experiment, also different shapes have to be expected. As shown in previous work \cite{rupp2012}, about 10-20\% of all clusters do not reach spherical shape during coagulation-driven growth but freeze out in non-spherical twin structures. We found three events of double imaging of twin clusters. One example is shown in \fref{fig:fig3}\,a). The patterns are matched by simple simulations of a twin particle using the PyScatman simulation tool \cite{colombo2022}, underlining that the same twin cluster was imaged from opposing directions.

In addition the size analysis of all pump-probe hits at 0\,ps showing spherical clusters supports our assumption. We determined all cluster radii via $R = 0.61 \cdot \lambda / \sin \Theta_{\rm 1stMin}$ with wavelength $\lambda$ and scattering angle of the first minimum $\Theta_{\rm 1stMin}$. At 0\,ps we can clearly confirm almost identical sizes in 87\,\% of all pattern pairs.\footnote{For six CDI pattern pairs the first minimum is indistinct or lost in the central hole, but the radial profiles match.} In \fref{fig:fig3}\,b) an example for such a twice imaged spherical xenon cluster is shown, while \fref{fig:fig4}\,a) presents a comparison of cluster sizes of 20 randomly selected pattern pairs at 0\,ps (out of the 87\,\%).

For the remaining pattern pairs, either largely differing cluster sizes were imaged, indeed stemming most probably from two different clusters, or the situation is inconclusive, e.g.\@ because of an insufficient quality of one pattern. 

At 0\,ps time-delay, for most pattern pairs with pump pattern and probe pattern from the same spherical cluster, an actual slight difference in cluster radius is observed, as in \fref{fig:fig3}\,b). On average, for all size-matching pattern pairs, the difference in cluster radius is 3.3\,nm (at maximum 9\,nm). As will be discussed in the following section, this effect can be attributed to a varying cluster position for every event. The position variations are indeed the strongest limitation to the time resolution of the XMOV approach in its current realization.\footnote{The distance between a particle and the detector also affects the achievable spatial resolution, but this is of minor importance here.}

\section{Temporal resolution analysis}
\label{sec:sec4}

\begin{figure}[tb]
\centering
\includegraphics[width=1.0\textwidth]{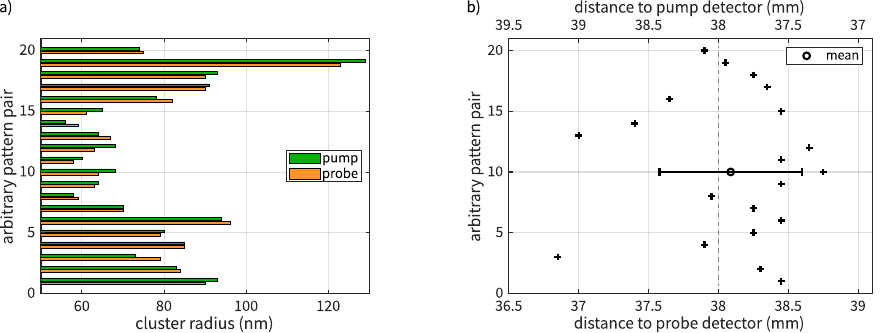}
\caption{Analysis of CDI pattern pairs at 0\,ps. a) Comparison of the cluster sizes of 20 randomly selected pattern pairs out of the 87\,\% pattern pairs showing the same single cluster. The slight size difference originates from assuming the clusters are exactly in the middle between both detectors. b) Actual variation of the cluster position between the CDI detectors. Distances to the respective detectors are retrieved for the same twenty randomly selected pattern pairs at 0\,ps time-delay. The mean value is shifted to the designed center position of 38\,mm (vertical dashed line) by about 0.1\,mm and has a standard deviation of about $\pm 0.6$\,mm, indicated as horizontal bar.}
\label{fig:fig4}
\end{figure}

The slight difference in cluster radius at 0\,ps time-delay of on average of 3.3\,nm is a result of the assumption that the clusters are exactly equidistant from the two detectors, when calculating the cluster size. In reality, however, the cluster positions are fluctuating along the beam propagation axis, caused by the width of the cluster jet. We can turn this around and assume instead that for a zero-delay hit, the size of the cluster has to be the same for both patterns. This allows us to retrieve the actual cluster position between the two detectors, by matching the radial profiles of pump hit and probe hit by varying detector distances, taking into account they have a fixed distance to each other. In \fref{fig:fig4}\,b), the optimized cluster positions between the detectors are plotted for some arbitrarily selected events (same as in figure 4\,a)) with both pump and probe hit. We find a standard deviation of the cluster position of $\pm0.6$\,mm from a slightly shifted center position of 0.1\,mm towards the pump detector. This width of the interaction region matches approximately the geometrically calculated full cluster jet width of 1.4\,mm, given by the nozzle distance and the skimmer size and position. 

The fluctuating cluster positions result in a reduced time resolution of the XMOV setup. If the cluster is one millimeter closer to the probe detector, i.e.\@ at 37.1\,mm instead of 38.1\,mm distance, it will be irradiated not at the set time-delay of the split-and-delay unit, indeed it will be irradiated 3.35\,ps later by the probe pulse and 3.35\,ps earlier by the pump pulse. In consequence, the $\pm0.6$\,mm translate into a statistic time-delay error of 8\,ps. 

To this statistical error, a systematic time-delay uncertainty of 4\,ps from the split-and-delay unit at this time has to be added.\footnote{Meanwhile, DESC has been upgraded, allowing even for femtosecond stability \cite{sauppe2018}.} For the below discussed results at large time-delays, this uncertainty of in total $\sqrt{(8\,\textnormal{ps})^2 + (4\,\textnormal{ps})^2} = 9$\,ps is of minor importance. Concepts to improve the time resolution are described in \sref{sec:sec6}.

\section{Analysis of pattern pairs with time-delays}
\label{sec:sec5}

With the knowledge from 0\,ps time-delay that the same cluster is imaged twice in the majority of pattern pairs, the time-resolved pattern pairs can now be examined. Because of the low quality of the patterns and the pre-processing required, the information that can be gained from the patterns is limited to the size and approximate shape of the cluster. Also, the statistics are too low to follow the dynamics in a sequence as conceptually described earlier. Instead, we investigate two phenomena observed for lower ($I \approx 2 \cdot 10^{12}$\,W/cm$^2$, $\tau = 100$\,fs) and higher ($I \approx 4 \cdot 10^{12}$\,W/cm$^2$, $\tau = 200$\,fs) pump flux.\footnote{Even if the standard deviation of the FEL pulse energy specified in \tref{tab:tab2} is considered, the pump fluxes differ significantly.} Note, for the cluster size calculation the average distances to each detector from \fref{fig:fig4}\,b) are used.

\subsection{Localized cluster expansion}
\label{sec:subsec2}

\begin{figure}[tb]
\centering
\includegraphics[width=1.00\textwidth]{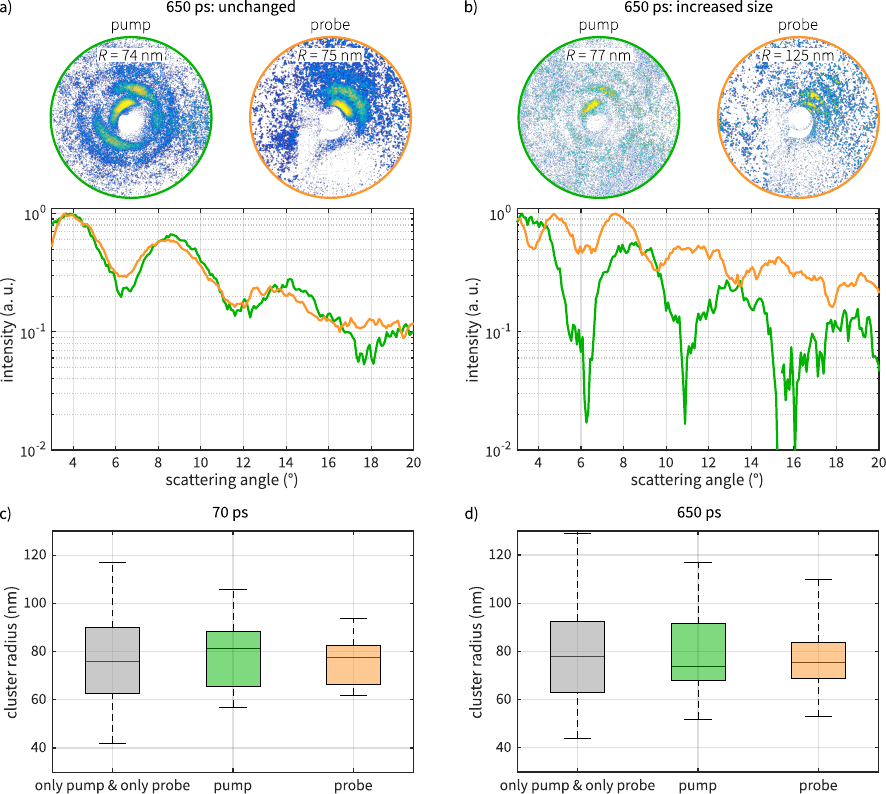}
\caption{CDI pattern pairs and their radial profiles at 650\,ps, shown up to a scattering angle of $20^{\circ}$. a) Initial spherical cluster with unchanged probed size. b) Initial spherical cluster with increased probed size or different probed cluster. c) and d) Boxplots (central horizontal line = median, box = values between the lower 25\,\% and upper 75\,\%, whiskers = min. and max. values) for comparison of the pump and probe cluster sizes for the time-resolved CDI pattern pairs with unchanged cluster size (as in a)). Additionally patterns recorded only with the pump pulse and only with the probe pulse are shown (the respective other beam was blocked).}
\label{fig:fig5}
\end{figure}

\begin{figure}[tb]
\centering
\includegraphics[width=1.00\textwidth]{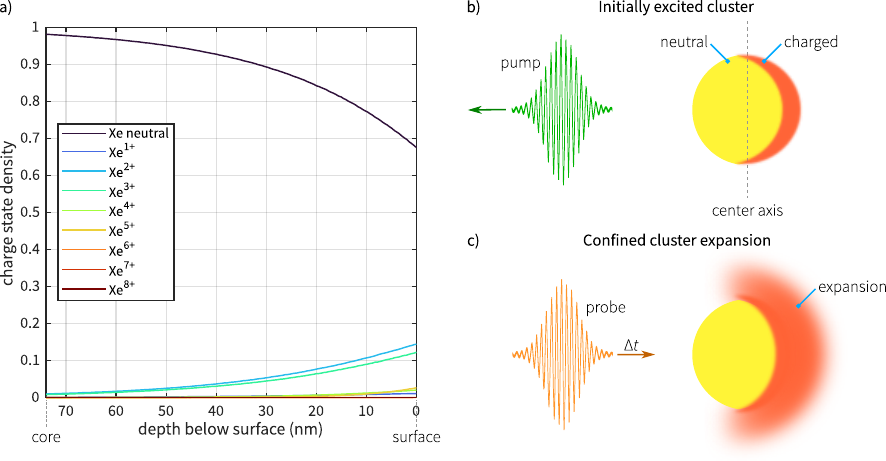}
\caption{Simulation and schematics (sectional view) of confined cluster ionization and subsequent partial expansion. a) For the pump pulse intensity in the first experiment and 74\,nm cluster radius (as in \fref{fig:fig5}\,a)) an one-dimensional Monte Carlo simulation reveals only a minority of atoms is charged, and only weakly. On the cluster surface a mean charge state of 0.9 is calculated. b) Illustration of the only partly penetrating charge into the cluster. For a 74\,nm cluster radius (as in \fref{fig:fig5}\,a)) this penetration leads to a crescent shape. c) The probe pulse facing cluster side and size remains virtually unchanged. Eventually the excited volume will alter, expand after some time-delay and finally leave behind a lentil-like shaped cluster after further time-delay.}
\label{fig:fig6}
\end{figure}

For xenon clusters heated with an optical laser, previous studies showed a fast surface softening which is reflected in vanishing diffraction signal starting at higher scattering angles \cite{peltz2014,gorkhover2016,peltz2022}, and a full fragmentation on longer timescales leading to speckles in the CDI patterns corresponding to density fluctuations \cite{fluckiger2016}. 

One may speculate that a surface softening could lead to a lower detection probability for the probe patterns and interpret the increasing occurrence of patterns with time-delay in the pump-probe configuration, which only show a pump pattern, as a sign of surface softening. At 0\,ps the ratio of pump patterns to probe patterns is $97 / 443 \approx$ 1:4.6 and reduces for 70\,ps to 1:2.6 and for 650\,ps to 1:1.2 (\tref{tab:tab3}). However, due to the low statistics, these numbers should be viewed with caution.

An unambiguous observation on the other hand is the missing structural change of clusters up to longest timescales, when pumped with $I \approx 2 \cdot 10^{12}$\,W/cm$^2$. Apparently, even after 650\,ps the majority of the observed clusters are still mostly intact, like the example shown in \fref{fig:fig5}\,a). For 70\,ps only in one probe pattern an increased cluster radius of 22\,nm and for 650\,ps in four probe patterns a mean increased radius of 46.8\,nm was found. An example of an increased cluster size at 650\,ps is shown in \fref{fig:fig5}\,b). However, comparing their occurrance with the statistics at zero time-delay (\sref{sec:sec3}), it seems also likely that in these cases a second nearby cluster was imaged by the probe pulse instead of an expanded cluster. Analysing all CDI pattern pairs of spherical shapes with a pump hit and a probe hit at 70\,ps and 650\,ps for the cluster sizes, we find both the median and the distributions of sizes unchanged, i.e.\@ within the range of the clusters position fluctuation (\sref{sec:sec4}). \Fref{fig:fig5}\,c) and d) show the direct comparison of the cluster size distributions at 70 and 650 ps. In absolute numbers an unchanged size is found for 84\,\% at 70\,ps and for 70\,\% at 650\,ps.

However, when a pump pattern is recorded, inevitably energy is deposited into the cluster, initiating dynamics and forcing an expansion and disintegration. How is it possible, that the only Van-der-Waals bound systems with a few meV binding energy remain intact after being sufficiently irradiated to cast a detectable diffraction pattern?

To explain this observation, we propose a nonuniform cluster expansion, which is confined to the cluster side facing the pump pulse. This idea is based on a simple estimation of the pump pulse induced excitation. For neutral xenon at 91.8\,eV photon energy, a short absorption length\footnote{Defined as incident intensity drop to $1/ \rme$ \cite{attwood2017}.} of $l_{\rm abs} = 1 / n_{\rm a} \sigma (\lambda) = 24$\,nm is calculated from a number density of solid xenon of $n_{\rm a} = 1.7 \cdot 10^{22}$\,cm$^{-3}$ and an absorption cross section of $\sigma =$ 25\,Mb. Also the lower charge states up to Xe$^{4+}$ exhibit high absorption cross sections at this photon energy \cite{andersen2001,emmons2005,aguilar2006}. An one-dimensional Monte Carlo simulation \cite{rupp2020} furthermore shows after the interaction with an FEL pulse with $I = 1.6 \cdot 10^{12}$\,W/cm$^2$, i.e.\@ when $1.9 \cdot 10^6$ photons fall into the geometrical cross section of a $R = 74$\,nm cluster, atoms at the surface have accumulated an average charge state of 0.9 (weighted mean). Already in a depth of $\approx 30$\,nm below the cluster surface, about 90\,\% of all atoms are still neutral, as shown in \fref{fig:fig6}\,a).

Hence, only the cluster side facing the pump pulse is ionized, illustrated as a crescent shape in \fref{fig:fig6}\,b). Continuing secondary ionization processes like electron impact ionization and relaxation processes as electron-ion recombination \cite{arbeiter2014} will alter the excited volume and make it more uniform, but the expansion may still be restricted to a finite volume as illustrated in \fref{fig:fig6}\,c). Due to the high absorption also for the probe pulse there is no possibility to detect any changes on the pump side, as long as they stay in the geometrical cross section of the cluster. Consequently, in the two-dimensional projection of the opposing probe pulse, the residual cluster has still the shape of a circle with virtually unchanged cluster size, as observed in the time-delay probe patterns. 

Assuming this mechanism, the final three-dimensional shape is expected to be comparable to a lentil. This cannot be directly retrieved from our data due to the low quality patterns. Moreover, no direct diffraction features of the expanding volume can be observed and the timescale of the expansion is an open question, but possibly enters the nanosecond time-regime \cite{fluckiger2016}. Also whether such a lentil shape further decays at even longer timescales remains subject to further studies.

\subsection{Cluster disintegration}
\label{sec:subsec3}

\begin{figure}[tb]
\centering
\includegraphics[width=1.00\textwidth]{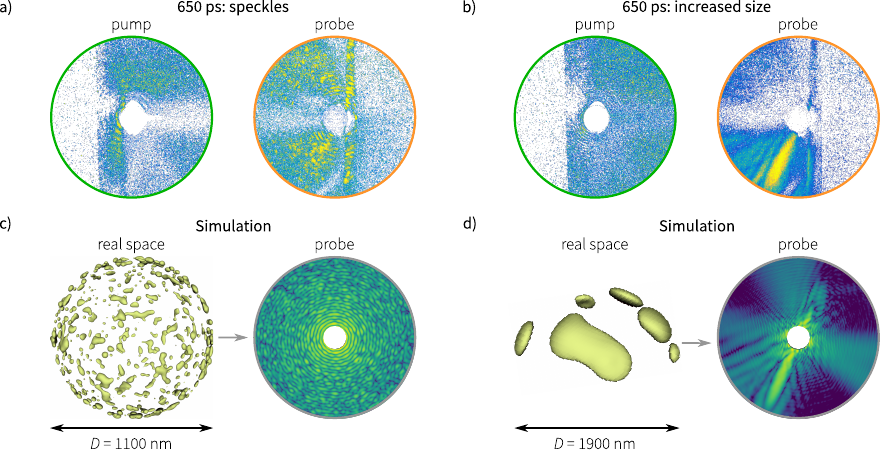}
\caption{CDI pattern pairs and exemplarily matching simulations at 650\,ps from the second experiment with higher focal intensities, shown up to a scattering angle of $20^{\circ}$. a) Initial spherical cluster which shows in the pump and probe pattern very narrow concentric rings and in the probe speckles at larger scattering angles. b) Initial very large spherical cluster with reflex-like structures in the probe pattern. Concentric rings are present but harder to spot because they are very fine. c) Simulated probe pattern to match a) using several fragments in real space, distributed around an empty inner area. d) Simulated probe pattern to match b) using only few large fragments in real space.}
\label{fig:fig7}
\end{figure}

In the second experiment, the higher focal intensities and longer pulse duration enabled us to make cluster disintegration visible. Both, the higher pump intensity helped to introduce more pronounced dynamics, and the higher probe intensity revealed more details. Although the patterns were still of limited quality, from 400\,ps on distinct speckles and reflex-like features could be observed. Two examples for 650\,ps are presented in \fref{fig:fig7}\,a) and b). Simulations of distributions reproducing these basic features are given in \fref{fig:fig7}\,c) and d), produced with the PyScatman simulation tool \cite{colombo2022}. For these simulations we assumed that the fragmenting clusters would break into smaller or larger fragments which expand radially, leaving the inner part empty. We note that these model shapes represent only guesses, and also other spatial distributions could explain the observed features. The real-space objects cannot be inferred from the CDI patterns due to the strong stray light contamination and detector non-linearity precluding reconstruction or forward fitting. The model however in principle agrees with previous results from infrared laser heated rare gas clusters \cite{fluckiger2016} that suggested an expanding cloud of gas and fragments with a certain density fluctuation nanoseconds after laser irradiation.

\section{Possibilities for advancing XMOV concept}
\label{sec:sec6}

\begin{figure}[tb]
\centering
\includegraphics[width=1\textwidth]{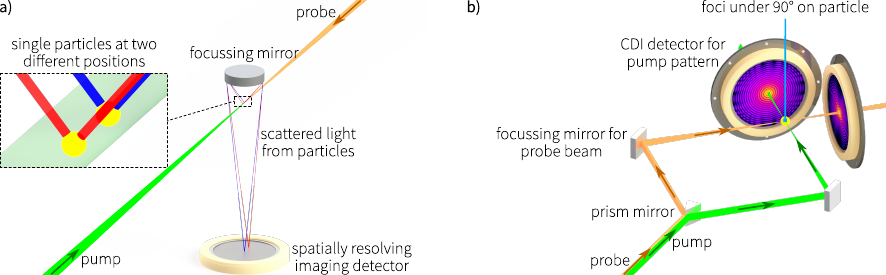}
\caption{Proposals for advancements of the XMOV. a) The particle's position along the focused beams can be mapped by focussing perpendicularly scattered light or the particle's fluorescence on a perpendicular spatial resolving detector. With a resolution of 1\,\textmu m, ten times magnified on the detector, the time resolution could be improved to 7\,fs. b) To image ultrafast excitation in three dimensions, in this concept pump and probe beam are focused perpendicular onto the sample. The pump pulse anisotropic excitation is imprinted into the probe pulse produced CDI pattern.}
\label{fig:fig8}
\end{figure}

To exploit the full potential of the experimental method higher focal intensities and an improved time resolution would be required. A time resolution below 100\,fs could be achieved with a two-dimensional mapping of the cluster's position along the focused beams with an additional perpendicular position sensitive detector, either detecting a single cluster's fluorescence or perpendicularly scattered light. A possible geometry is shown in \fref{fig:fig8}\,a). Additionally, this detector would indicate if one or more clusters are within the focal volumes and hence allow to reject multiple hits.

For a deeper investigation of the pump-pulse-induced dynamics we propose an adapted setup with both beam paths intersecting the particle perpendicularly, see \fref{fig:fig8}\,b). In this concept, the pump-pulse-induced dynamics, confined only to the side of the cluster facing the pump pulse, is directly imprinted into the CDI probe pattern. We note, that a similar approach was already proposed several years ago \cite{schmidt2008}. This geometry would allow not only to image the partial cluster expansion and the possibly final shape, but with an increased time resolution and shorter FEL pulses \cite{milne2017,serkez2018,coffee2019,maroju2021,trebushinin2023} also resolve nanoplasma development penetrating into the particle \cite{bostedt2012,rupp2020}.
In this proposal, a \mbox{(near-)} perpendicular incidence angle would allow to image simultaneously the ``front'' and ``back'' of the cluster with the probe pulse. However, depending on the experimental conditions, a compromise towards smaller angles might be necessary and could be even beneficial as grazing incidence would also give a higher flexibility in wavelength.

\section{Summary}
\label{sec:sec7}

We developed a new experimental method to study X-ray induced structural dynamics in nanoparticles with coherent diffractive imaging. The particularity of the method is the imaging of the initial particle and its induced dynamics in two non-overlying and time-resolved CDI patterns. The proof of concept is demonstrated in two experiments on xenon clusters with same-color XUV pump-probe pulses. For a lower pump focal intensity and a maximum time-delay of 650\,ps, still intact clusters are observed. We interpret them as a result of an expansion confined only to the cluster side facing the pump pulse, leaving behind lentil-like shaped residuals. For higher focal intensities, in contrast, from about 400\,ps time-delay on, a full disintegration is observed.
Our first realization and demonstration of a single-particle X-ray movie camera shows the challenges but also the vast potential of pump-CDI probe-CDI approaches to make X-ray induced changes on the nanoscale visible.

\ack

The success of the highly complex experiments within the short period of a beamtime was only possible due to the tremendous effort of many people. We gratefully thank the TUB IOAP workshop and the DESY FS-Verbundwerkstatt, Frank Scholze \& Christian Laubis and the team of PTB's EUV radiometry laboratory for the mirror's reflectivities characterization, Ralf Steinkopf and Johannes Stock for help with focussing mirrors. We acknowledge DESY (Hamburg, Germany), a member of the Helmholtz Association HGF, for the provision of experimental facilities. The experiments were carried out at FLASH at the CAMP endstation at beamline BL1. Beamtime was allocated for proposals F-20140076 and F-20160533. We acknowledge the Max Planck Society for funding the development and the initial operation of the CAMP end-station within the Max Planck Advanced Study Group at CFEL and for providing this equipment for CAMP@FLASH. The installation of CAMP@FLASH was partially funded by BMBF grants 05K10KT2, 05K13KT2, 05K16KT3 and 05K10KTB from FSP-302. D.Ro. was supported by the Chemical Sciences, Geosciences, and Biosciences Division, Office of Basic Energy Sciences, Office of Science, US Department of Energy, Grant No. DEFG02-86ER13491, and, earlier, by the Helmholtz Gemeinschaft through the Young Investigator Program. Further funding is acknowledged from DFG via grants Mo 719/13 and Mo 719/14 and from
Leibniz-Gemeinschaft via Grant No. SAW/2017/MBI4. M.S., A.C., and D.R. received further funding from Swiss National Science Foundation under Grant No. 200021E-193642.

\section*{Data availability}
The discussed experimental data are available at \url{https://zenodo.org/doi/10.5281/zenodo.10391943}. The full raw data sets are available on request.

\section*{References}
\bibliography{XMOV_Ref}

\end{document}